\documentstyle[twocolumn]{jpsj}
\input epsf.tex

\def\runtitle{Single Impurity Anderson Model with Coulomb Repulsion 
between Conduction Electrons 
on the Nearest-Neighbour Ligand Orbital}
\def\runauthor{Ryu {\sc Takayama} and Osamu {\sc Sakai}}

\title
{
Single Impurity Anderson Model with Coulomb Repulsion 
between Conduction Electrons 
on the Nearest-Neighbour Ligand Orbital 
}

\author
{ 
Ryu {\sc Takayama}\footnote{E-mail address: takayama@cmpt01.phys.tohoku.ac.jp}$^{1,2}$ 
and Osamu {\sc Sakai}$^{1}$
}

\inst
{
$ ^{1}$Department of Physics, Tohoku University, Sendai 980-8578 \\
$ ^{2}$Research Institute of Electrical Communication, Tohoku University, Sendai 980-8577\\
}

\recdate
{
February 25, 1998
}

\abst
{
We study how the Kondo effect is affected by the Coulomb interaction 
between conduction electrons on the basis of a simplified model. 
The single impurity Anderson model is extended to include 
the Coulomb interaction on the nearest-neighbour ligand orbital. 
The excitation spectra are calculated using the numerical 
renormalization group method.
The effective bandwidth on the ligand orbital, $D^{\rm eff}$, 
is defined to classify the state. 
This quantity decreases as the Coulomb interaction increases. 
In the $D^{\rm eff} > \Delta$ region, 
the low energy properties are described by the Kondo state, 
where $\Delta$ is the hybridization width. 
As $D^{\rm eff}$ decreases in this region, 
the Kondo temperature $T_{\rm K}$ is enhanced, and its magnitude 
becomes comparable to $\Delta$ for $D^{\rm eff} \sim \Delta$. 
In the $D^{\rm eff} < \Delta$ region, 
the local singlet state between the electrons 
on the $f$ and ligand orbitals is formed.
}

\kword
{
Coulomb repulsion between conduction electrons,
numerical renormalization group method,
magnetic excitation,
single particle excitation, 
charge excitation
}

\begin{document}
\sloppy
\maketitle

%
%\section{Introduction}
%\label{sec:intro}
%

Recently, much attention has been paid to the Kondo effect 
for systems with Coulomb 
interaction between the conduction electrons (CCE).
\cite{igarashi95,khaliullin95,schork94,schork96,davidovich97,tornow97} 
It was reported that a small CCE enhances the Kondo temperature $T_{\rm K}$. 
A correlated host has usually been given by the Hubbard model, 
and the studies have been carried out by means of the 
perturbation theory on the CCE because its influence 
on the Kondo effect can be treated only by such procedures. 
Therefore it is not clear how the CCE affects the magnetic 
impurity problem when it is not weak. 
Detailed properties of the low energy excitation are also 
unknown, and thus it is not clear whether low energy 
properties can be described as the Kondo state. 
In this letter we restrict ourselves to a simplified 
model but study the effect of CCE in detail. 
We choose a model which includes the Coulomb interaction between conduction 
electrons only on the nearest-neighbour site of the magnetic impurity. 
By this simplification, change in the bulk host states caused 
by CCE is excluded from the consideration in the strict sense. 
But the effects through the local change around the impurity can be studied 
in detail. 
We expect that gross features caused by the CCE can be extracted 
from the studies of the present model.\cite{inf-Hubbard0} 
We calculate the dynamical excitation spectra such as the 
single particle and the magnetic excitation without restricting ourselves to the weak 
CCE cases.
From these studies we show that the low energy properties are 
well described as the usual Kondo state with increased effective $T_{\rm K}$ 
when the effective bandwidth of the conduction electron, $D^{\rm eff}$, 
is larger than the hybridization width of the $f$-electron, $\Delta$. 
As the strength of CCE increases, $D^{\rm eff}$ gradually decreases. 
When $D^{\rm eff}$ becomes comparable to $\Delta$, 
the effective $T_{\rm K}$ also becomes comparable to $\Delta$. 
The excitation spectra change their characteristics rapidly when 
the strength of the CCE increases further. 

%
%\section{Model Hamiltonian}
%\label{sec:model}
%

In this letter we present the calculation of the dynamical excitation spectra for 
the following model Hamiltonian, 
\begin{eqnarray}
H & = & H_{\rm A} + H_{\rm L},  \label{eq:2.1} \\
H_{\rm A} & = & H_{\rm f} + H_{\rm hyb} + H_{\rm c},  \label{eq:2.2} \\
H_{\rm f} & = & \varepsilon_{\rm f} \sum_\sigma n_{\rm f \sigma} + \frac{U_{\rm ff}}{2} \sum_{\sigma
 \neq \sigma'} n_{\rm f \sigma} n_{\rm f \sigma'}, \\
H_{\rm hyb} & = & \frac{V}{\sqrt{N}} \sum_{{\bf k},\sigma} (f_\sigma^{+} c_{{\bf k} \sigma} + h.c.), \\
H_{\rm c} & = & \sum_{{\bf k},\sigma}  \varepsilon_{\bf k} c_{{\bf k} \sigma}^{+} c_{{\bf k} \sigma}, \\
H_{\rm L} & = & \varepsilon_{\rm L} \sum_\sigma n_{\rm L \sigma} + \frac{U_{\rm L}}{2} \sum_{\sigma
 \neq \sigma'} n_{\rm L \sigma} n_{\rm L \sigma'}. \label{eq:hp}
\end{eqnarray}
Equation (\ref{eq:2.2}) represents the 
single impurity Anderson model (SIA) using the standard notation. 
Energies $\varepsilon_{\rm f}$ and $\varepsilon_{\bf k}$ 
are measured relative to the Fermi level, $E_{\rm F}$.
The Coulomb interaction between electrons on the ligand orbital 
is given by the term $H_{\rm L}$.
The operator $n_{\rm L \sigma}$ is the occupation number on the ligand orbital 
on the nearest-neighbour site, and is  defined as 
$n_{\rm L \sigma}=c_{\rm L \sigma}^{+} c_{\rm L \sigma}$ with 
$c_{\rm L \sigma}=(1/\sqrt{N})\sum_{\bf k} c_{{\bf k} \sigma}$. 
Here the quantities $\varepsilon_{\rm L}$ and $U_{\rm L}$ are 
the energies of the orbital and the Coulomb interaction on it, 
respectively.
We restrict ourselves to the electron-hole symmetric case, i.e., 
$2\varepsilon_{\rm f} + U_{\rm ff} =0$ and $2\varepsilon_{\rm L} + U_{\rm L} =0$.
Hereafter, we parameterize the hybridization intensity, $V$, 
by the hybridization width, $\Delta=\pi V^2/2D$.
In this letter, we use the numerical renormalization group (NRG) method 
to calculate the single particle and the magnetic excitation spectra. 
All results are obtained using the discretization parameter 
$\Lambda=1.5$.
\cite{sakai92} 
%
%\section{Results and Discussion}
%
%
\begin{fullfigure}[htbp]
\begin{center}
\epsfxsize=17cm \epsfbox{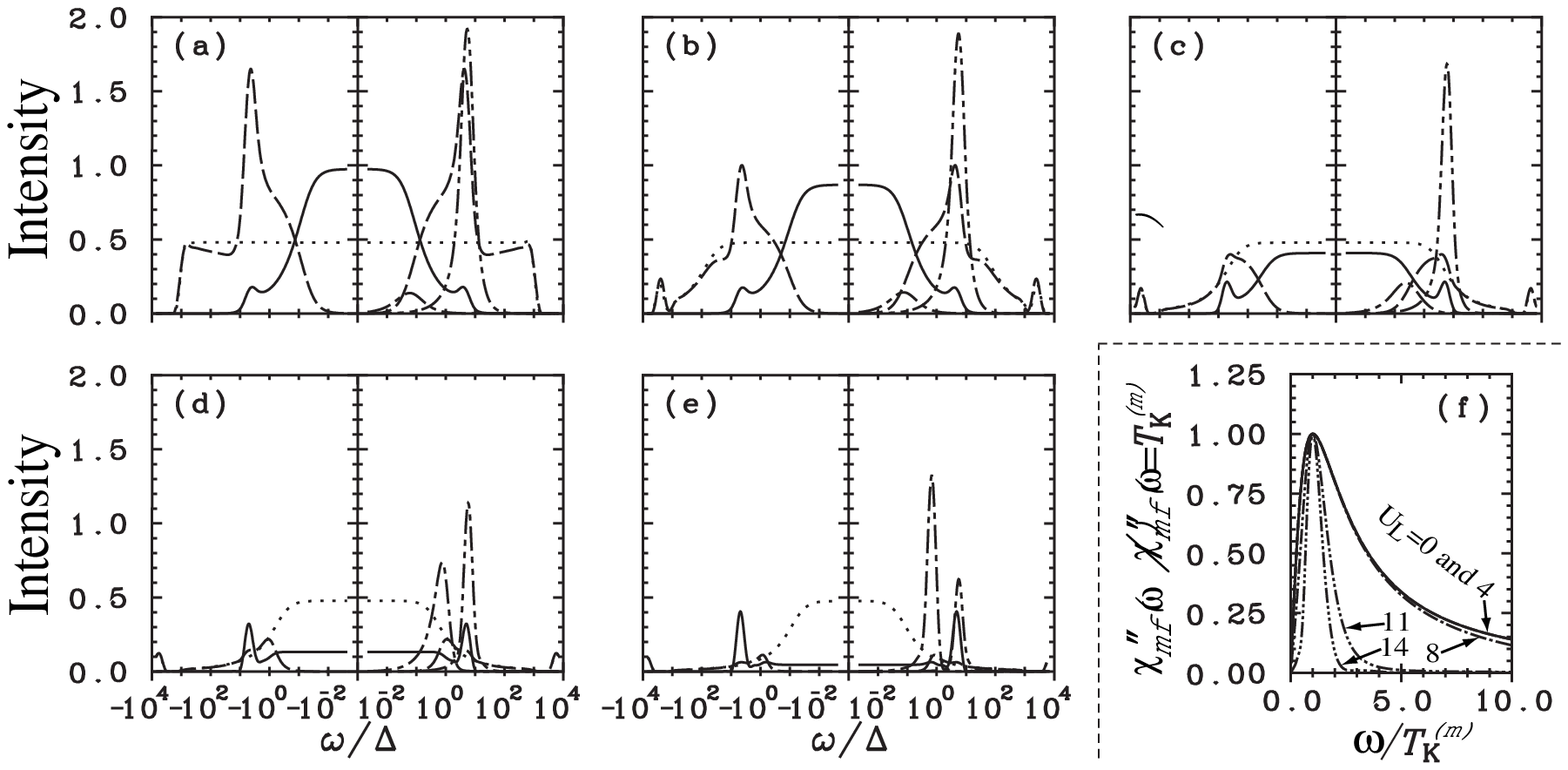}
%\epsfile{file=./fig1.eps,scale=0.87}
\end{center}
\caption{
The excitation spectra 
for various $U_{\rm L}$ values 
in the cases of $U_{\rm ff}/D=8.0 \times 10^{-3}$ 
and $\Delta/D=3.0 \pi \times 10^{-4}$.
From (a) to (e), 
the abscissa is the excitation energy and ranges from $10^{-3}$ to $10^{4}$ 
on the logarithmic scale in units of hybridization $\Delta$. 
The value of $U_{\rm L}$ is 0.0~(a), 
4.0~(b), 8.0~(c), 11.0~(d) and 14.0~(e).
In these figures, 
the lines show 
the single particle excitation spectra of $f$-electron  
$\pi \Delta \rho_{\rm f}/N_{\rm f}$ (solid line)
and that of $c$-electron on the ligand orbital $\rho_{\rm L}/N_{\rm f}$ (broken line),
magnetic excitation spectra $T_{\rm K}^{\rm (m)} \chi_{\rm mf}''/N_{\rm f} \times 10.0$ 
(one-dot-dash line) and 
charge excitation spectra $\chi_{\rm cf}''/N_{\rm f} \times 0.2$ (two-dot-dash line).
Here $N_{\rm f}$ is the degeneracy factor, $N_{\rm f}=2$.
The DOS of the conduction electrons for $H_{\rm L}+H_{\rm c}$ model, 
$\rho_{\rm L0}/N_{\rm f}$, 
is also plotted as a dotted line on the same scale. 
Figure~(f) shows $\chi_{\rm mf}''$ on the 
linear energy scale in units $T_{\rm K}^{\rm (m)}$.
}
\label{fig:spec8}
\end{fullfigure}
We consider a band with a constant density of states 
which extends from $-D$ to $D$ for simplicity.
\cite{inf-Hubbard}

Figure~\ref{fig:spec8} shows the excitation spectra for various 
cases of $U_{\rm L}$. 
In these figures, parameters $U_{\rm ff}$ and $\Delta$ are fixed at 
$U_{\rm ff}/D=8.0 \times 10^{-3}$ and 
$\Delta/D=3.0 \pi \times 10^{-4}$.\cite{parameters}
For $U_{\rm L}=0$, this parameter set gives the Kondo regime 
of SIA, $U_{\rm ff} \gg \Delta$.

%%% \Delta=0 model

First, we examine the $U_{\rm L}$ dependence of $\rho_{\rm L0}$, 
which is the density of states (DOS) on the nearest-ligand orbital 
for a fictitious model setting as $\Delta=0$. 
It is indicated by the dotted lines. 
For $\Delta=0$, the Hamiltonian (\ref{eq:2.1}) is decomposed into two 
independent parts, $H_{\rm f}$ and $H_{\rm L}+H_{\rm c}$. 
The part $H_{\rm L}+H_{\rm c}$ corresponds to the Wolff model of the magnetic impurity.
The quantity $\rho_{\rm L0}$ is normalized to have an integrated intensity equal to 1. 
Therefore, the intensity at $\omega=0$ should be 0.5 according to 
Friedel's sum rule.
As seen from Fig.~\ref{fig:spec8}(a) the present calculation 
fulfills this condition with about $97\%$  accuracy, and seems to be 
sufficient for obtaining qualitative conclusions.

\begin{figure}[htbp]
\begin{center}
\epsfxsize=8cm \epsfbox{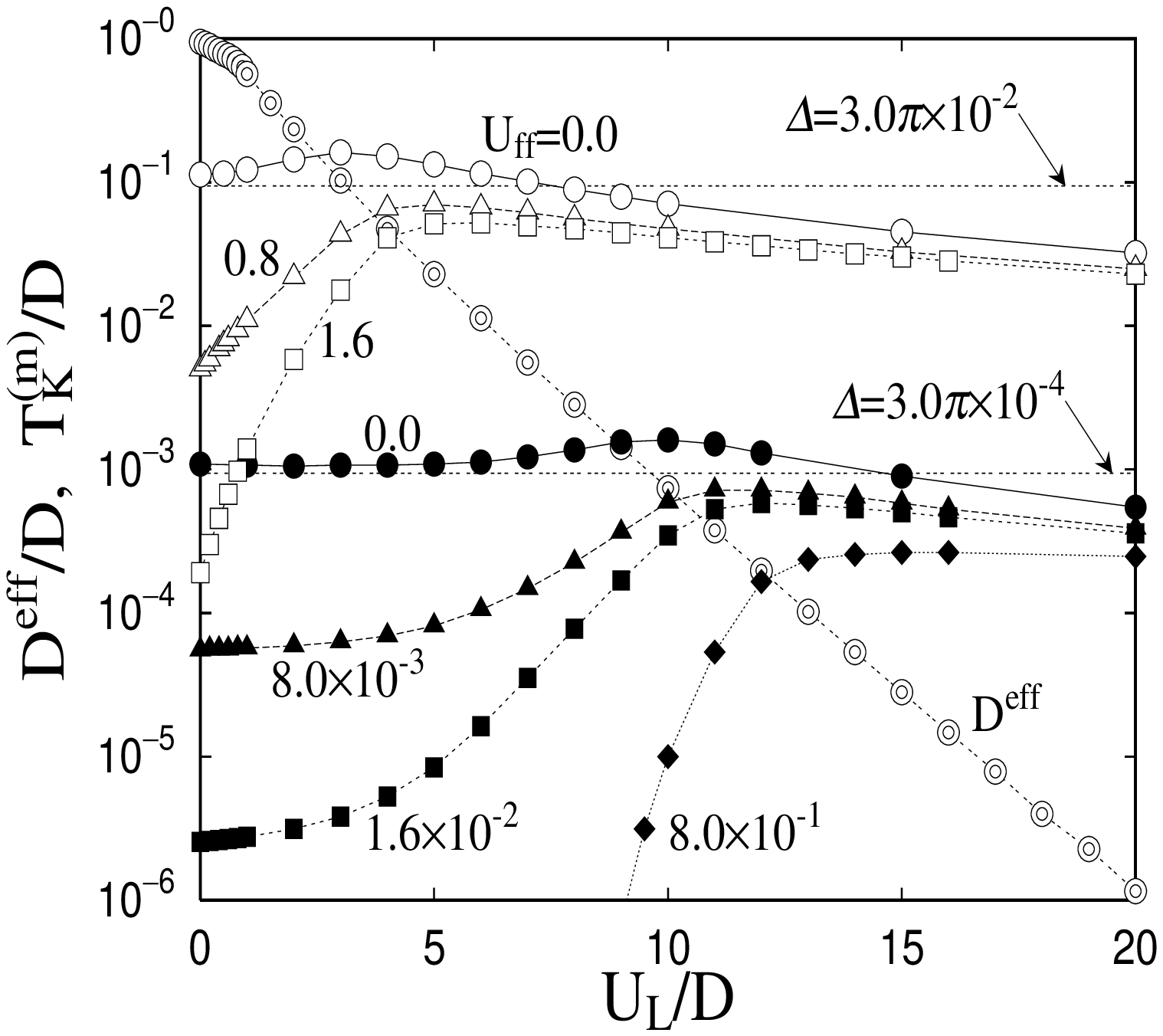}
%\epsfile{file=./fig2.eps,scale=0.50}
\end{center}
\caption{
The Kondo temperature defined as the energy of the peak position of the magnetic
excitation spectra of the $f$-electron, $T_{\rm K}^{\rm (m)}/D$, 
and the effective bandwidth defined as the energy of the peak position of the magnetic 
excitation spectra of the $c$-electron for the $H_{\rm L}+H_{\rm c}$ model,
$D^{\rm eff}/D$, as a function of $U_{\rm L}$/D.  The double circles
indicate $D^{\rm eff}/D$, and the other open symbols denote
$T_{\rm K}^{\rm (m)}/D$ for $\Delta/D=3.0\pi\times10^{-2}$ cases.  The solid symbols
indicate $T_{\rm K}^{\rm (m)}/D$ for $\Delta/D=3.0\pi\times10^{-4}$ cases.
Curves are a guide to the eye.
}
\label{fig:tk-upp}
\end{figure}

When $U_{\rm L}=0$, $\rho_{\rm L0}$ is given by the rectangular DOS
which extends from $-D$ to $D$ as shown in Fig.~\ref{fig:spec8}(a).
As seen from Fig.~\ref{fig:spec8}(b), 
$\rho_{\rm L0}$ shows a three-peaked-structure for the case $U_{\rm L}/2 > D$. 
The peak which appears in the energy range 
$|\omega/\Delta|<10^{2}$ corresponds to the Kondo resonance of the $c$-electron 
for the $H_{\rm L}+H_{\rm c}$ model.
Furthermore, satellite peaks are also found at $\omega/\Delta \sim \pm 2.0 \times 10^{3}$. 
These energies correspond to the atomic excitation, 
$\omega/D \sim \pm U_{\rm L}/2$. 
As $U_{\rm L}$ increases, 
the width of the Kondo resonance of $c$-electron  becomes narrow. 
This may be regarded as the reduction of the effective bandwidth 
caused by the CCE.
We define the effective bandwidth, $D^{\rm eff}$, as the peak position
of the magnetic excitation of the $H_{\rm L}+H_{\rm c}$ model.
This quantity almost coincides with the half-width 
of the Kondo peak of $\rho_{\rm L0}$. 
We have the relation $D^{\rm eff}=0.951D (\equiv D_{0})$ for $U_{\rm L}=0$. 
The quantity $D^{\rm eff}$ is shown by the double 
circles in Fig.~\ref{fig:tk-upp} as a function of $U_{\rm L}$. 

%%%

%%% Kondo temperature

Next, we see how the magnetic excitation spectra 
of the $f$-electron, $\chi_{\rm mf}''$, depend on $U_{\rm L}$ for $\Delta \neq 0$.
In Fig.~\ref{fig:spec8}, $\chi_{\rm mf}''$ is plotted as a one-dot-dash line. 
Its intensity is obtained by multiplying the energy of 
its peak position, $T_{\rm K}^{\rm (m)}$.
We adopt this value, $T_{\rm K}^{\rm (m)}$, 
as the characteristic energy of the low-energy excitation,
because it is lower than that of other excitations such as charge fluctuation. 
We note here that $T_{\rm K}^{\rm (m)}$ has a value of about $0.6 T_{\rm K}$
for the conventional SIA in the Kondo regime, 
where $T_{\rm K}$ denotes the Kondo temperature determined from the magnetic
susceptibility at $T=0$, $\chi$, as $T_{\rm K}=1/4\chi$.

In Fig.~\ref{fig:spec8}(a), $T_{\rm K}^{\rm (m)}$ is at 
$\omega/\Delta =5.87 \times 10^{-2}$.
As $U_{\rm L}$ increases, 
$T_{\rm K}^{\rm (m)}$ initially shifts to the high-energy side (from Fig.~(a) to (d)) 
and returns to the low-energy side (from (d) to (e)). 
We note that the maximum value of $T_{\rm K}^{\rm (m)}$ has 
magnitude comparable to $\Delta$ ((d)). 

We replot the magnetic excitation spectra in Fig.~\ref{fig:spec8}(f). 
The energy is scaled by $T_{\rm K}^{\rm (m)}$, 
and the spectra are normalized by the peak heights. 
The spectra almost coincide up to values of $U_{\rm L}/D \leq 8$. 
This means that the magnetic excitation 
has an almost universal shape, 
and therefore the low energy properties 
can be described as the Kondo state characterized by the energy scale 
$T_{\rm K}^{\rm (m)}$. 
As mentioned later, this feature can be seen when the condition 
$D^{\rm eff} > \Delta$ holds.
On the other hand,  
the line shape of $\chi_{\rm mf}''$ rapidly changes to narrow 
at around $D^{\rm eff} \sim \Delta$ when $U_{\rm L}$ increases further.

%%%

The quantity, $T_{\rm K}^{\rm (m)}$, as a function of $U_{\rm L}/D$
for various $U_{\rm ff}$ cases 
is shown in Fig.~\ref{fig:tk-upp}.
It increases as $U_{\rm L}$ increases 
in the $D^{\rm eff} > \Delta$ region, 
while it 
decreases gradually in the $D^{\rm eff} < \Delta$ region.
The magnitude of the maximum value is comparable to $\Delta$ 
both in the extreme Kondo regime ($U_{\rm ff} \gg \Delta $) 
shown by the solid diamond symbols
and the valence fluctuation (VF) regime ($U_{\rm ff} \ll \Delta $) 
shown by the solid circle symbols.
In the extreme large $U_{\rm L}$ region, i.e.,
$D^{\rm eff} \ll \Delta$ region,
$T_{\rm K}^{\rm (m)}$ decreases as 
$T_{\rm K}^{\rm (m)}/D \propto 16\Delta/\pi (U_{\rm L}+U_{\rm ff})$. 

In the previous studies, 
the increase of $T_{\rm K}^{\rm (m)}$ in the weak CCE region is ascribed to the 
renormalization of the Kondo exchange interaction, $J$. 
\cite{khaliullin95,schork96,tornow97}
It is modified from $J$ to $J(1+\gamma)$ with the 
renormalization factor, $\gamma \propto U_{\rm L}/D$, 
when one uses the perturbation theory 
in the lowest order of the Coulomb interaction. 
To check the relation, we examine the $U_{\rm L}$ dependence of $T_{\rm K}^{\rm (m)}$. 
The modification of $J$ leads to the relative change of $T_{\rm K}^{\rm (m)}$ 
\cite{khaliullin95}
\begin{equation}
\frac{T_{\rm K}^{\rm (m)}}{T_{\rm K0}^{\rm (m)}}
= \exp\left\{\frac{\pi U_{\rm ff}}{8 \Delta}\frac{\gamma}
{1+\gamma}\right\}. \label{eq:smallu}
\end{equation}
Here we use $J=8\Delta/\pi U_{\rm ff}$ for the symmetric SIA and 
$T_{\rm K0}^{\rm (m)}$ is the value of $T_{\rm K}^{\rm (m)}$ for $U_{\rm L}=0$.
The quantity $\gamma$ is obtained from eq.~(\ref{eq:smallu})
\begin{equation}
\gamma = \left[1-\frac{8\Delta}{\pi U_{\rm ff}}
\log\left(\frac{T_{\rm K}^{\rm (m)}}{T_{\rm K0}^{\rm (m)}}\right)\right]^{-1} -1 .
\label{eq:gamma}
\end{equation}
Using the results shown in Fig.~\ref{fig:tk-upp}, 
we found that the linear relation $\gamma= \alpha U_{\rm L}/D$ 
holds well in the range $0\leq U_{\rm L}/D \leq 0.6$.
Here $\alpha$ itself is a function of $\Delta$ and $U_{\rm ff}$. 
The value of $\alpha$ is given as 
$(U_{\rm ff}/D, \alpha)=
 (0.8, 2.6 \times 10^{-1})$ and 
$(1.6, 3.6 \times 10^{-1})$ 
for $\Delta/D=3.0 \pi \times 10^{-2}$, 
$(8.0 \times 10^{-3}, 5.4 \times 10^{-3})$ and 
$(1.6 \times 10^{-2}, 8.8 \times 10^{-3})$ for $\Delta/D=3.0 \pi \times 10^{-4}$.

%%%%%

\begin{figure}[htbp]
\begin{center}
\epsfxsize=8cm \epsfbox{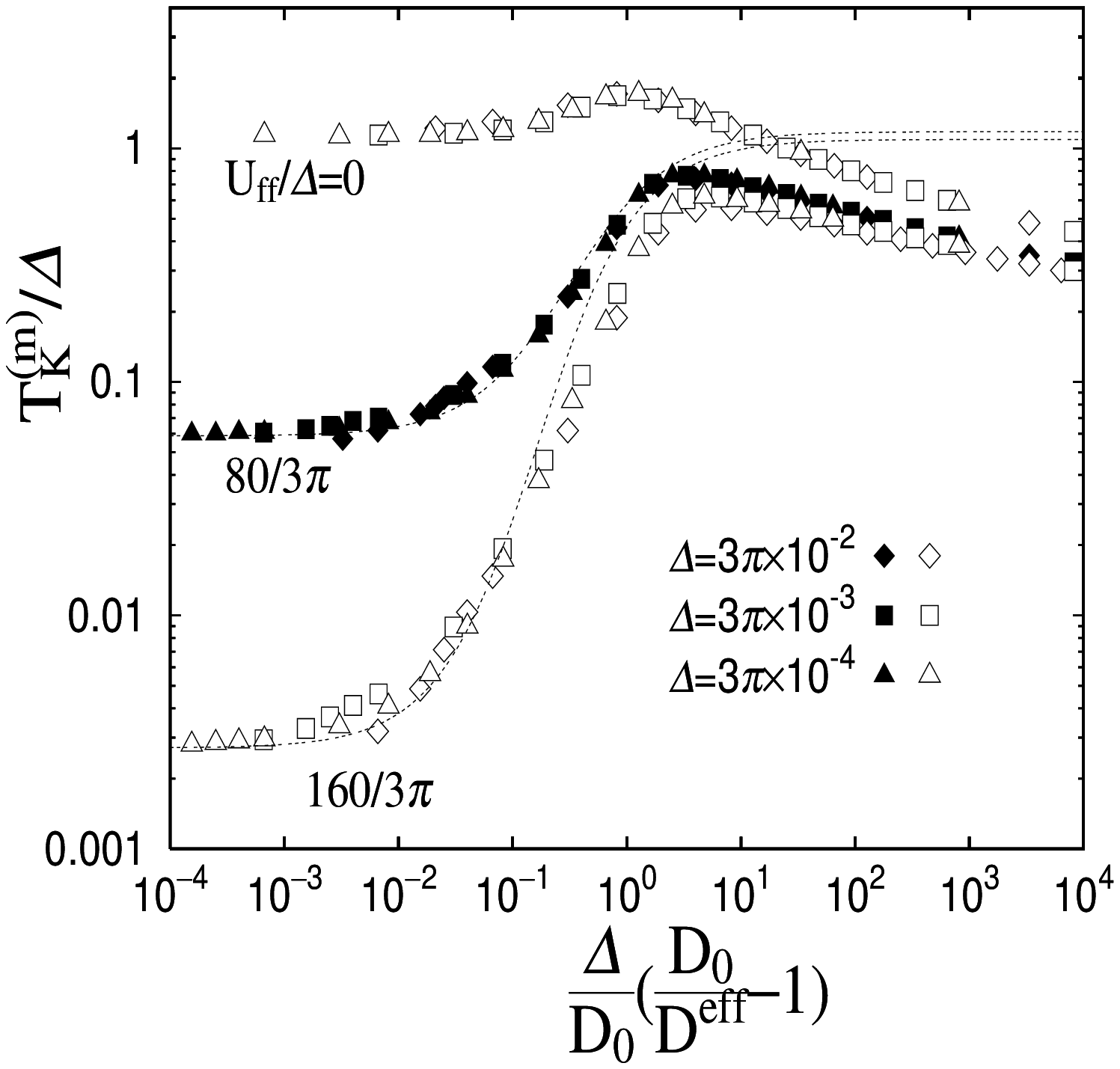}
%\epsfile{file=./fig3.eps,scale=0.50}
\end{center}
\caption{
The Kondo temperature defined as the energy of the peak position of the magnetic
excitation spectra, $T_{\rm K}^{\rm (m)}/\Delta$,
as a function of $(\Delta/D_{0})\cdot(D_{0}/D^{\rm eff}-1)$
for three values of $U_{\rm ff}/\Delta$.
The dotted lines show the fitted results using 
$\gamma'=\alpha'(\Delta/D_{0})(D_{0}/D^{\rm eff}-1)$ and eq.~(7). 
See the text for detail. 
Here $\alpha'=3.2~(U_{\rm ff}/\Delta=80/3\pi)$ and $6.0~(160/3\pi)$.
}
\label{fig:tk-sc}
\end{figure}

In order to examine the larger $U_{\rm L}$ region, 
we replot the results of Fig.~\ref{fig:tk-upp} 
in terms of the relation 
$T_{\rm K}^{\rm (m)}/\Delta$ vs $(\Delta/D_{0})\cdot(D_{0}/D^{\rm eff}-1)$ 
as shown in Fig.~\ref{fig:tk-sc}. 
Here $D_{0}$ is the value of $D^{\rm eff}$ for $U_{\rm L}=0$.
The quantity $D_{0}/D^{\rm eff}$ corresponds to the enhancement factor of 
the specific heat of the $c$-electron for $H_{\rm L}+H_{\rm c}$ model. 
If $D^{\rm eff}$ is expressed as $D^{\rm eff}=D_{0} \exp(-\pi U_{\rm L}/4D)$, 
the quantity $(D_{0}/D^{\rm eff}-1)$ is proportional 
to $U_{\rm L}/D$ for small $U_{\rm L}/D$. 
When the value of $U_{\rm ff}/\Delta$ is fixed, the data points 
for various $\Delta$ cases are almost on a common curve.
We try to reproduce the curves by using the quantity 
$\gamma' = \alpha'(\Delta/D_{0})(D_{0}/D^{\rm eff}-1)$
and substitute it for $\gamma$ in eq.~(\ref{eq:smallu}), 
where $\alpha'$ is the fitting parameter.  
The enhancement factor $\gamma$ in eq.~(\ref{eq:gamma}) 
is proportional to $(\Delta/D_{0})(D_{0}/D^{\rm eff}-1)$
in $(\Delta/D_{0})(D_{0}/D^{\rm eff}-1)<0.1$ region. 
We determine the value of $\alpha'$ within this region.
The fitted results are plotted by dotted lines in Fig.~\ref{fig:tk-sc}, 
and they reproduce $T_{\rm K}^{\rm (m)}$ qualitatively
even in the region $D^{\rm eff} \sim \Delta$.

Let us return to Fig.~\ref{fig:spec8}.
The solid and the two-dot-dash lines are the single particle excitation spectra (SPE) of
the $f$-electron, $\rho_{\rm f}$, and the charge excitation spectra, $\chi_{\rm cf}''$, 
respectively.
The intensity of $\rho_{\rm f}$ is illustrated by multiplying 
the factor $\pi\Delta/N_{\rm f}$.
Therefore, the intensity at $\omega=0$ should be 1.0 
in Fig.~\ref{fig:spec8}(a) for the case of $U_{\rm L}=0$. 
The present calculation fulfills this condition
with about 97\% accuracy. 
In Fig.~\ref{fig:spec8}(a), 
the Kondo resonance appears in the energy region $|\omega/\Delta| < 10^{-1}$, 
and the atomic-like excitation appears at $\omega/\Delta \sim \pm 4.0$ 
which correspond to $\omega \sim \pm U_{\rm ff}/2$.
The peak position of $\chi_{\rm cf}''$ almost coincides with that of the 
atomic-like excitation. 
The DOS on the nearest-ligand orbital, $\rho_{\rm L}$, 
show a depression in the energy region $|\omega/\Delta| < 10^{-1}$ 
as shown by the broken line. 
It behaves as $\rho_{\rm L} \propto \omega^2$ in this region. 
Two peaks appear on each side of the depressed region. 
They have shoulder and peak structures which are caused 
by the Kondo resonance at $\omega/\Delta \sim \pm 10^{-1}$ and 
by the atomic excitation at $\omega \sim \pm U_{\rm ff}/2$, 
respectively.

When $U_{\rm L}$ increases, the width of the Kondo resonance 
increases in the $D^{\rm eff} > \Delta$ region as seen from 
Figs.~\ref{fig:spec8}(b) and \ref{fig:spec8}(c). 
This increase can be ascribed to the enhancement of the effective 
exchange coupling, and is consistent with the increase of $T_{\rm K}^{\rm (m)}$. 
The intensity of $\rho_{\rm f}$ at $\omega=0$ gradually decreases 
when $U_{\rm L}$ increases.\cite{upp0} 
These facts indicate that the hybridization coupling between $f$ and 
ligand orbital is enhanced by the Coulomb interaction.
The atomic excitation peak is shifted towards higher energy by a small amount 
and the spectral shape becomes narrow. 

In the case of $D^{\rm eff} < \Delta$, the magnetic excitation has a rather 
sharp peak at the energy defined as $T_{\rm K}^{\rm (m)}$. 
The intensity of $\rho_{\rm f}$ in the energy region 
$|\omega| < T_{\rm K}^{\rm (m)}$ is very small as shown in Fig.~\ref{fig:spec8}(e). 
On the other hand, $\rho_{\rm f}$ and $\rho_{\rm L}$ have small 
peaks at about $|\omega| \sim T_{\rm K}^{\rm (m)}$. 
The energy of the atomic excitation is not significantly changed. 
The excitation energy $T_{\rm K}^{\rm (m)}$ is roughly given by 
$16\Delta/\pi (U_{\rm L}+U_{\rm ff})$, as already noted. 
This energy corresponds to the exchange coupling between the 
electrons on the $f$ and the nearest-ligand orbital. 
These facts may indicate that the $f$ state and its nearest-neighbour 
state form a local spin singlet pair when the charge 
fluctuation is strongly suppressed by the $U_{\rm L}$ term. 
This local singlet pair couples weakly to band states in the outer region. 
As noted previously, $\chi_{\rm mf}''$ rapidly changes its line shape 
around the cross-over region 
from the Kondo singlet region ($D^{\rm eff} > \Delta$ region)
to the local singlet one ($D^{\rm eff} < \Delta$ region).

We note that the cross-over 
can also be seen from the analysis of the flow chart of 
the renormalized energy level~ (FCEL) in the NRG. 
\cite{km}
We have analyzed the FCEL of the present model, 
and found that the fixed point is the local Fermi
liquid state in both cases,  
but the difference lies in the asymptotic area to the fixed point.
In the $D^{\rm eff} > \Delta$ case, 
the low energy levels are similar to that of the 
conventional SIA, i.e., 
they are classified as the ordinary 
Kondo state. 
However, in the $D^{\rm eff} < \Delta$ case,
the energy levels indicate that the local spin singlet pair 
and the conduction electrons are almost independent 
of each other.\cite{phase}

%
%\section{Summary}
%

In summary, we have examined the impurity Anderson model with CCE 
on the nearest-ligand orbital. 
The interaction $U_{\rm L}$ reduces the effective bandwidth 
for the $c$-$f$ hybridization, $D^{\rm eff}$. 
The characteristics of the system depend on whether 
the effective bandwidth  
$D^{\rm eff}$ is larger or smaller than $\Delta$. 
In the $D^{\rm eff} > \Delta$ region, 
the characteristic energy of the magnetic excitation, 
which is defined as $T_{\rm K}^{\rm (m)}$, increases until its 
magnitude becomes comparable to $\Delta$ when $U_{\rm L}$ increases. 
The low energy properties are expected to be given as the Kondo state. 
The increase of $T_{\rm K}^{\rm (m)}$ is approximately expressed as 
the enhancement of the hybridization. 
When $U_{\rm L}$ increases further into the 
$D^{\rm eff} < \Delta$ region, the magnetic excitation energy decreases 
as $16\Delta/\pi (U_{\rm L}+U_{\rm ff})$, 
reflecting the formation of the spin singlet pair state 
from the electrons on the $f$-orbital 
and its nearest-neighbour orbital.

The present model includes CCE only on the ligand orbital to which 
the $f$-orbital has the direct hopping matrix. 
\cite{df-coulomb} 
However, we expect that the present results will be generalized 
to the model with CCE for the band states, 
if one concentrate on the effects due to the modification of the 
hybridization process. 
The quantity $D^{\rm eff}$ of the present model should be 
interpreted as the effective bandwidth of such a band CCE model. 
The calculation including the self-consistent procedure 
based on the $d=\infty$ model to extract 
a true effective bandwidth 
will be attempted in the near future.\cite{sakai94,georges96}

%
%\section*{Acknowledgment}
%

The authors thank Dr.~S.~Suzuki and Mr.~W.~Izumida for useful discussions.
We also thank Dr.~H.~Kusunose for valuable comments.
One of the authors (R.~T.) was partly supported by the Kasuya scholarship.
The present work is supported by Grants-in-Aid No.~06244104, No.~09244202,  
and No.~09640451 from the Ministry of Education, Science and Culture.


\begin{thebibliography}{99}
\label{ref}

\bibitem{schork94} T.~Schork and P.~Fulde:
Phys.~Rev.~B{\bf 50} (1994) 1345.

\bibitem{igarashi95} J.~Igarashi, K.~Murayama and P.~Fulde:
Phys.~Rev.~B{\bf 52} (1995) 15966.

\bibitem{khaliullin95} G.~Khaliullin and P.~Fulde:
Phys.~Rev.~B{\bf 52} (1995) 9514.

\bibitem{schork96} T.~Schork:
Phys.~Rev.~B{\bf 53} (1996) 5626.

\bibitem{davidovich97} B.~Davidovich and V.~Zevin: preprint (cond-mat/9706283).

\bibitem{tornow97} S.~Tornow, V.~Zevin and G.~Zwicknagl: preprint(cond-mat/9701137).

\bibitem{inf-Hubbard0}
For example, the Hubbard model in the infinite dimensional space ($d=\infty$ model) 
can be mapped onto the impurity Anderson model. 

\bibitem{sakai92}  O.~Sakai, Y.~Shimizu and T.~Kasuya: Prog.~Theor.~Phys.~Suppl. {\bf 108} (1992) 73.

\bibitem{inf-Hubbard}
When we consider the $d=\infty$ Hubbard model, 
the effective bandwidth for the mapped impurity Anderson model 
is reduced strongly by the Coulomb interaction.
But in this letter we set the bandwidth, $2D$, to be a constant, 
independent of $U_{\rm L}$. 
However we note that as shown later the width of the single 
particle excitation spectra on the ligand site is 
greatly reduced by $U_{\rm L}$ similar to the effective 
bandwidth of the $d=\infty$ model.

\bibitem{parameters} 
In the usual Anderson model, the low-energy scale 
(for example the Kondo temperature) depends strongly on the ratio 
$U_{\rm ff}/\Delta$ , but does not depend strongly on $U_{\rm ff}/D$ 
as far as $U_{\rm ff}/D < 1$. 
So we set $U_{\rm ff}/D \ll 1$ to remove the additional 
dependence on $U_{\rm ff}/D$. 
In the strong $U_{\rm L}$ cases, $U_{\rm L}/D \gg 1$, 
we have the relation $U_{\rm L} \gg U_{\rm ff}$ in the present model. 
In actual cases the relation $U_{\rm ff} > U_{\rm L}$ should be expected, 
but we expect that the essential point will be unchanged when the parameters are 
arranged by the ratio $U_{\rm ff}/\Delta$ and $U_{\rm L}/D$.

\bibitem{upp0} 
The decrease of $\rho_{\rm f}$ at $E_{\rm F}$ is ascribed to 
the cross-effect due to the $U_{\rm ff}$ and $U_{\rm L}$ terms. 
Indeed, when we put $U_{\rm ff}=0$, the intensity of $\rho_{\rm f}$ at $E_{\rm F}$ 
is unchanged. 

\bibitem{km}  H.~B.~Krishna-murthy, J.~W.~Wilkins and K.~G.~Wilson: Phys.~Rev. B~{\bf 21} (1980) 1003; {\it ibid.} 1044.

\bibitem{phase} We note that the phase shift of the conduction electron is 
also $\pi/2$ in this case. 

\bibitem{df-coulomb} In realistic situations, the Coulomb interaction 
between the $f$-electron and the conduction electrons is also 
important, see for example, R.~Takayama and O.~Sakai: 
J.~Phys.~Soc.~Jpn. {\bf 66} (1997) 1512.

\bibitem{georges96} A.~Georges, G.~Kotliar, W.~Krauth and M.~J.~Rozenberg:
Rev.~Mod.~Phys. {\bf 68} (1996) 13 and references therein.

\bibitem{sakai94} O.~Sakai and Y.~Kuramoto:
Solid State Commun. {\bf 89} (1994) 307.

\end{thebibliography}
\end{document}